\documentclass[twoside]{article}
\usepackage[utf8]{inputenc}
\usepackage{amsmath}
\usepackage{amsfonts}
\usepackage{amssymb}
\usepackage{mathtools}
\usepackage[labelformat=simple]{subcaption}
\usepackage{bm}
\usepackage{graphicx}
\usepackage{algorithm} 
\usepackage{algpseudocode}
\usepackage[numbers,sort]{natbib}
\usepackage{color}
\usepackage{hyperref}
\usepackage{bm}
\usepackage{url}
\usepackage{amsthm}
\usepackage{todonotes}
\usepackage{lipsum}
\usepackage{physics}
\usepackage{array}
\usepackage{stackengine}
\usepackage{geometry}
\newcommand\barbelow[1]{\stackunder[1.2pt]{$#1$}{\rule{.8ex}{.075ex}}}
\newcommand{\diagentry}[1]{\mathmakebox[1.8em]{#1}}
\newcommand{\xddots}{%
  \raise 4pt \hbox {.}
  \mkern 6mu
  \raise 1pt \hbox {.}
  \mkern 6mu
  \raise -2pt \hbox {.}
}

\usepackage{enumitem}

\title{Contributed Discussion of \\``A Bayesian Conjugate Gradient Method"}
\author{F-X. Briol$^{1,2}$, F. A. DiazDelaO$^{3}$, P. O. Hristov$^{3}$\\ $^{1}$University College London, $^{2}$The Alan Turing Institute \\ $^{3}$Institute for Risk and Uncertainty, University of Liverpool}

\begin{document}

\maketitle
		We would like to congratulate the authors of \cite{Cockayne2019} on their insightful paper, and welcome this publication which we firmly believe will become a fundamental contribution to the growing field of probabilistic numerical methods and in particular the sub-field of Bayesian numerical methods. In this short piece, we first initiate a discussion on the choice of priors for solving linear systems, then propose an extension of the Bayesian conjugate gradient (BayesCG) algorithm for solving several related linear systems simultaneously.
		
	\subsection*{Prior specification for Bayesian inference of linear systems}

		In the Bayesian paradigm, once a particular observation model is agreed upon, most of the work goes into selection of the prior. In the case of a linear system $A u = b$ and in particularly for conjugate gradient methods, our observation model consists of projections of $b$ observed without noise. The authors of \cite{Cockayne2019} place a Gaussian prior on $u$, which provides advantages to placing a prior on the inverse of the matrix $A$ \cite{Hen15}, including invariance to preconditioners. We agree that this is a significant advantage, but also think one could go much further in eliciting priors for solving linear systems, as is done for other Bayesian numerical methods. 

		In Bayesian quadrature, the task is to estimate $\Pi[f] = \int_{\mathcal{X}} f(x) \pi(x) \mathrm{d}x$, given evaluations of the integrand $f$ at some locations on the domain $\mathcal{X}$. Clearly, the quantity of interest is 
		$\Pi[f]$; yet, it is common to put a prior on $f$ instead, which then induces a prior on $\Pi[f]$. For differential equations, the problem is to find the solution $u$ of a system of equations $\mathcal{A} u(x) =g(x)$ (where $\mathcal{A}$ is some known integro-differential operator), given evaluations of $g$; and existing Bayesian methods also propose to specify a prior on $g$ instead of the quantity of interest $u$.
		In both cases, the main motivation for placing priors on latent quantities is that this is more natural, or convenient, from a modelling point of view. At the same time, it is often possible to inspect the mathematical expression for the latent quantity, or we may at least have some additional information about it, such as smoothness or periodicity information. In such cases, encoding this information in the prior leads to algorithms with fast convergence rates and tighter credible intervals, as demonstrated for these Bayesian integration and differential equation methods \cite{Cockayne2016,Briol2019PI}. We believe that the same is likely to be true for the case of linear systems.

		Indeed, in many applications, it is possible to know properties of $A$ beforehand, such as information on its spectrum, conditioning or sparsity. We argue that it is more natural to encode this knowledge in a prior, and it may in fact lead to a better calibration of uncertainty. To illustrate this, consider some of the systems of differential equations used in engineering to describe fluid flow and structural response to loading, which are usually discretised into a linear system.
		In computational structural mechanics the operator $\mathcal{A}$ can be used to describe the \textit{stiffness} of an assembled finite element model (FEM). Similarly, in computational fluid dynamics (CFD), $\mathcal{A}$ can represent mesh coefficient matrices. Since both of these matrices describe physical properties of the object under study, their sparsity patterns will be governed largely by the object's geometry. It is therefore common that analysts have some prior knowledge about $\mathcal{A}$, based on engineering insight and experience in solving similar systems.
		
\begin{figure}[h!]
			\centering
			\begin{subfigure}{0.28\textwidth}
				\includegraphics[width=\textwidth,trim={390 40 400 40},clip]{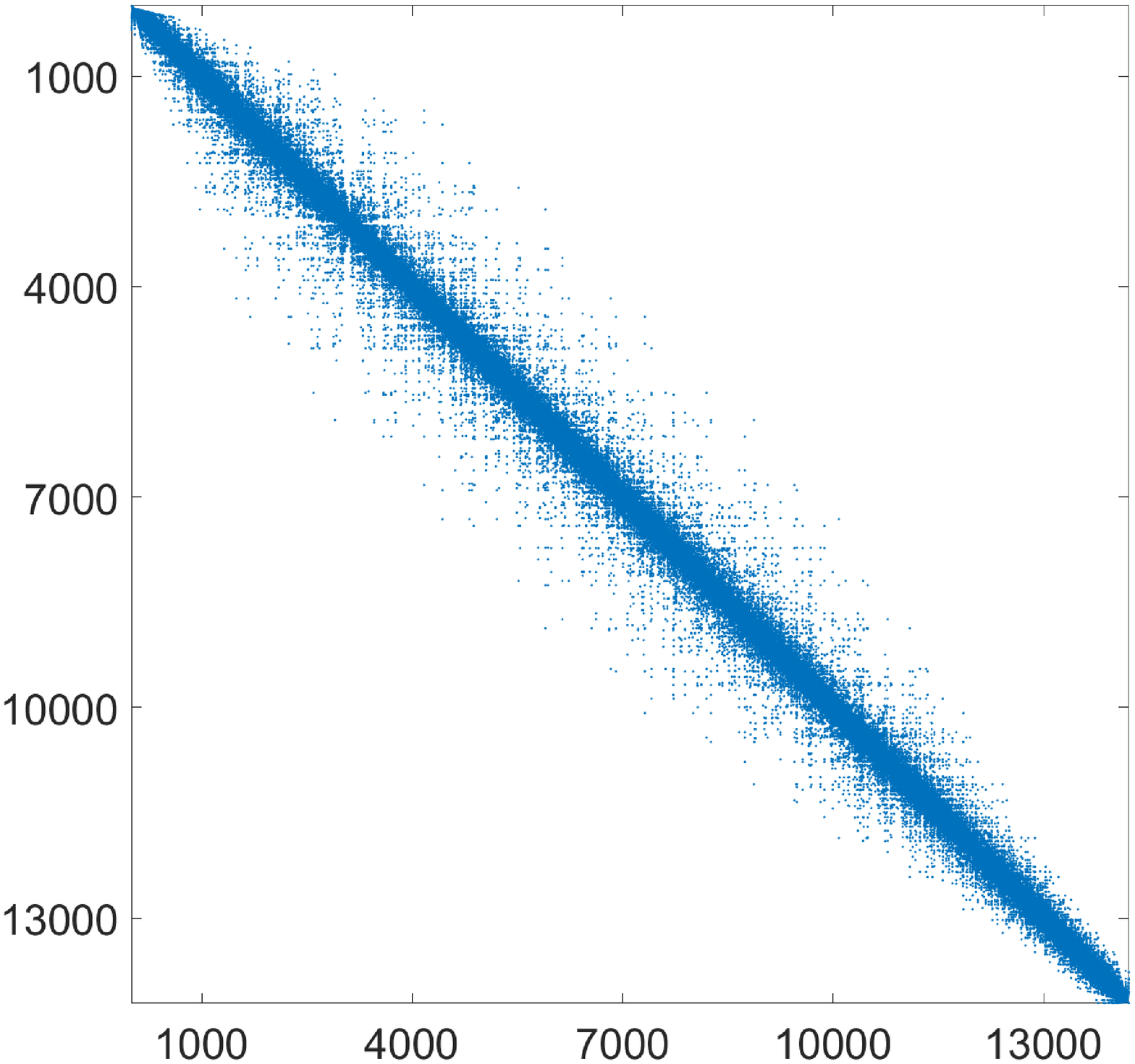}
				\caption{}
				\label{fig:mat_foil}
			\end{subfigure}
			\enskip
			\hspace{1cm}
			\begin{subfigure}{0.28\textwidth}
				\includegraphics[width=\textwidth,trim={390 40 400 40},clip]{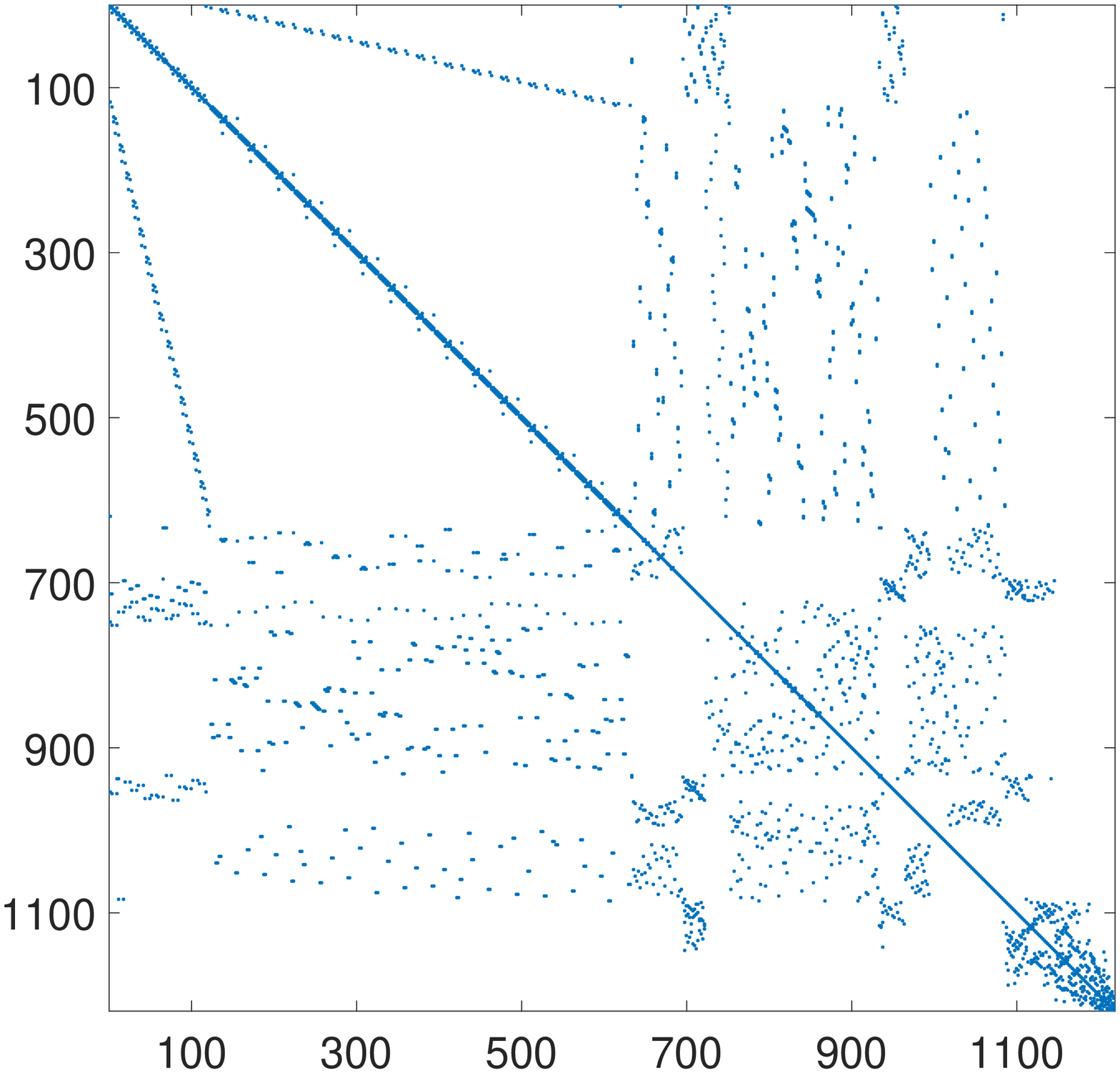}
				\caption{}
				\label{fig:mat_fan}
			\end{subfigure}
			\caption{Stiffness matrices with different degrees of sparsity and non-zero patterns. The systems described by these matrices are: 
			\subref{fig:mat_foil} a laminar airfoil; \subref{fig:mat_fan} jet engine compressor fan.}
			\label{fig:matrices}
		\end{figure}

		Figure \ref{fig:matrices} provides examples of the form of $A$ (i.e. discretisations of $\mathcal{A}$) for systems taking part in a typical coupled analysis of a jet engine compressor loading. The sparsity pattern shown in Figure~\ref{fig:mat_foil} encodes the coefficients of an unstructured mesh for a two dimensional airfoil in a CFD simulation \cite{Davis:2011}. The matrix in Figure~\ref{fig:mat_fan} depicts the FEM stiffness matrix of the compressor disc and blades. Both geometries were meshed with two-dimensional triangular elements. 
		In this context, the load on the compressor stage depends on the rotational speed and the force produced by its blades, which in turn depends on the rotational speed of the compressor. 
		Employing similar chains of coupled models is not uncommon in design and analysis of complex engineering systems, and can further complicate the choice of a prior model. We believe that eliciting such priors for coupled systems is a crucial question, very much aligned with one of the ambitions of probabilistic numerics: the propagation of uncertainty through pipelines of computation \cite{PN15}.


\subsection*{A generalisation to multiple linear systems}

BayesCG also provides an excellent opportunity to develop novel methodology for solving linear solvers. Suppose we have several linear systems which need to be solved either simultaneously or sequentially, such that for $j \in \{1,\ldots,J\}$, we want to solve\footnote{For simplicity of notation, we assume all systems are of the same size, but this could be generalised straightforwardly.}:
\begin{align*}
A_j x_j^* =  b_j,  
\end{align*}
where $A_j \in \mathbb{R}^{d \times d}$, $x_j^* \in \mathbb{R}^d$ and $b_j \in \mathbb{R}^d$ for some $d \in \mathbb{N}_{> 0}$. As discussed in 
\cite{DeRoos2017}, this is a common problem in statistics and machine learning. Take for example the issue of model selection for Gaussian processes: this includes calculating the log-marginal likelihood for several choices of covariance functions or covariance function hyperparameters, each requiring the solution of a linear system whose solutions will be closely related (atleast for similar choices of parameters). Similarly, for Bayesian inverse problems, the forward problem needs to be solved for several values of the parameters (perhaps over the path of some Markov chain Monte Carlo realisation), which will boil down to solving several closely related linear systems.

As principled Bayesians, it would be natural to construct a joint estimator on the solutions of these $J$ linear systems, rather than estimating the solutions independently. This is particularly the case if we know anything about how the solutions of these linear systems relate to one another, in which case information available through search directions in the $j^{\text{th}}$ system may be informative about the solution $x^*_{j'}$ for $j \neq j'$. This idea is closely related to transfer learning, which was recently advocated for problems in numerical analysis by \cite{Xi2018MultiOutput} (who focused on numerical integration). Although several methods exist to transfer information from one task to the other, such as recycled Krylov spaces \citep{DeRoos2017}, there are no existing Bayesian approach. 

Interestingly, we show below that the BayesCG algorithm of \cite{Cockayne2019} may be generalised straightforwardly to this setting. All expressions below are given so as to mirror the notation of the original algorithm. The main point to make is that all of these systems can be seen as a single, larger, linear system of the form $ \barbelow{A} \barbelow{x}^* = \barbelow{b}$ where $\barbelow{x} = ((x^*_1)^\top,\ldots,(x^*_J)^\top)^\top \in \mathbb{R}^{dJ}$, $\barbelow{b} = (b_1^\top,\ldots,b_J^\top)^\top \in \mathbb{R}^{dJ}$ and $\barbelow{A} \in \mathbb{R}^{dJ \times dJ}$ is of the form
\begin{align*}
	\barbelow{A} = \text{BlockDiag}\left[A_1,\ldots,A_J\right] = 
	\begin{pmatrix}
    \diagentry{A_1}\\
    &\diagentry{\xddots}\\
    &&\diagentry{A_J}\\
	\end{pmatrix}.
	\end{align*}
 We define the data obtained by $y_{i} = s_{i}^\top A x^* = s_{i}^\top b$ for $i \in \{1,\ldots,m\}$. We will define $\barbelow{S}_m \in \mathbb{R}^{dJ \times m}$ to be the matrix consisting of columns given by $m$ search directions. The data can therefore be expressed in vector form as $\barbelow{y}_m = \barbelow{S}_m^\top \barbelow{b}$. Taking a Bayesian approach, we select a prior of the form $\mathcal{N}(\barbelow{x},\barbelow{x}_0,\barbelow{\Sigma}_0)$, for some $\barbelow{x}_0 \in \mathbb{R}^{dJ}$ and $\barbelow{\Sigma}_0 \in \mathbb{R}^{dJ \times dJ}$. Conditioning on the data $\barbelow{y}_m$, we obtain a posterior of the form $\mathcal{N}(\barbelow{x} ; \barbelow{x}_m,\barbelow{\Sigma}_m)$ with $\barbelow{x}_m  = \barbelow{x}_0 + \barbelow{\Sigma}_0 \barbelow{A}^\top \barbelow{S}_m \barbelow{\Lambda}_m^{-1} \barbelow{S}_m^\top \barbelow{r}_0$, $\barbelow{\Sigma}_m  = \barbelow{\Sigma}_0 - \barbelow{\Sigma}_0 \barbelow{A}^\top \barbelow{S}_m \barbelow{\Lambda}_{m}^{-1} \barbelow{S}_m^\top \barbelow{A} \barbelow{\Sigma}_0$ where $\barbelow{r}_0 = \barbelow{b} - \barbelow{A} \barbelow{x}_0$ and $\barbelow{\Lambda_m} = \barbelow{S}_m^\top \barbelow{A} \barbelow{\Sigma}_0 \barbelow{A}^\top \barbelow{S}_m$. The search directions which allow us to avoid the matrix inverse are $\barbelow{A} \barbelow{\Sigma}_0 \barbelow{A}^\top$-orthogonal, and provide what we call the \emph{multi-system BayesCG algorithm}. Let $r_m = \barbelow{b} - \barbelow{A} \barbelow{x}_m$, $\barbelow{\tilde{s}}_1 = \barbelow{r}_0$ and $\barbelow{s}_m = \barbelow{\tilde{s}}_m/\|\barbelow{\tilde{s}}_m\|_{\barbelow{A} \barbelow{\Sigma}_0 \barbelow{A}^\top}$ for all $m$, then for $m>1$, assuming that $\tilde{\barbelow{s}}_m \neq \barbelow{0} = (0,\ldots,0)$, these directions are:
\begin{align*}
	\tilde{\barbelow{s}}_m & = \barbelow{r}_{m-1} - \langle \barbelow{s}_{m-1}, \barbelow{r}_{m-1} \rangle_{ \barbelow{A}\barbelow{\Sigma}_0 \barbelow{A}^\top} \barbelow{s}_{m-1}.
\end{align*}
At this point, most of the equations in the two paragraph above look identical to those in the paper, but include larger vectors and matrices. We now make several remarks:
\begin{enumerate}

	\item The search directions obtained through the multi-system BayesCG algorithm lead to some dependence across linear systems. That is, the estimator for $x_j^*$ for some fixed $j$ will be impacted by $A_{j'},b_{j'}$ for some $j' \neq j$. This dependence will come from the matrix $\barbelow{\Sigma}_0$, the covariance matrix of our prior. This leads to a larger computational cost, due to the fact that we are now having to perform matrix-vector products of matrices of size $dJ \times dJ$, but this may be acceptable if it provides improved accuracy and uncertainty quantification.

	\item Several special cases of prior matrix $\barbelow{\Sigma}_0$, inspired by vector-valued reproducing kernel Hilbert spaces or multi-output Gaussian processes, can be more convenient to use in practice due to their intepretability. One example are separable covariance functions, which were previously explored by \cite{Xi2018MultiOutput} for transfer learning in numerical integration. They take the form $\barbelow{\Sigma}_{0}= B \otimes \Sigma_{0}$ where $\otimes$ denotes the Kronecker product, $B \in \mathbb{R}^{J \times J}$ and $\Sigma_0 \in \mathbb{R}^{d \times d}$. In this case, the matrix $B$ can be seen as a covariance matrix across tasks (i.e. across linear systems), whilst $\Sigma_0$ is the covariance matrix which would otherwise be used for a single linear system. In particular, this approach would allow us to combined the algorithm with alternative transfer learning approaches, such as the Krylov subspace recycling discussed in \cite{DeRoos2017} which can be used to select $\Sigma_0$.

	\item In the case where $\barbelow{\Sigma}_0$ has block-diagonal form $\text{BlockDiag}\left[\Sigma_{0,1},\ldots,\Sigma_{0,J}\right]$ for $\Sigma_{0,1},\ldots,\Sigma_{0,J} \in \mathbb{R}^{d \times d}$, the multi-system Bayesian conjugate gradient method reduces to $J$ separate instances of the BayesCG; it is therefore a strict generalisation.

	\item The requirement that search directions are $\barbelow{A} \barbelow{\Sigma}_0 \barbelow{A}^\top-$orthogonal forces us to solve the $J$ linear systems simultaneously, obtaining one observation from each system at a given iteration of the multi-system BayesCG algorithm. This prevents us from considering the sequential case where we first solve $A_1$, then solve $A_2$ and so on. However, we envisage that alternative algorithms could be developed for this case, and could help provide informative priors in a sequential manner.

\end{enumerate}

\subsubsection*{Acknowledgments}

F-X. Briol was supported through the EPSRC grant [EP/R018413/1] and by The Alan Turing Institute's Data-Centric Engineering programme under the EPSRC grant [EP/N510129/1]. F. A. DiazDelaO acknowledges the support of The Alan Turing Institute, where he was a visiting fellow under the EPSRC grant [EP/S001476/1].

{

\small
\setlength{\bibsep}{0.5pt}
\bibliographystyle{plain}
\bibliography{peter_bib}

}

\end{document}